# Experimental Verification of Rate Flexibility and Probabilistic Shaping by 4D Signaling

**Fabian Steiner[(1)], Francesco Da Ros[(2)], Metodi Plamenov Yankov[(2)], Georg Böcherer[(1)], Patrick Schulte[(1)], Søren Forchhammer[(2)], Gerhard Kramer[(1)]**

[(1)] *Technical University of Munich, 80333 Munich, Germany*
[(2)] *DTU Fotonik, Technical University of Denmark, DK-2800 Kongens Lyngby, Denmark*
*fabian.steiner@tum.de*

**Abstract:** The rate flexibility and probabilistic shaping gain of 4-dimensional signaling is experimentally tested for short-reach, unrepeated transmission. A rate granularity of 0.5 bits/QAM symbol is achieved with a distribution matcher based on a simple look-up table.

**OCIS codes:** 060.2330, 060.1660, 060.4080.

## 1. Introduction

Flexibility is a key requirement for next-generation optical transponders to meet the demands of different deployment scenarios. Recently, probabilistic amplitude shaping (PAS) [1] gained broad interest as a means to realize shaping gains and a fine granularity in spectral efficiencies (SEs) for an extended reach and higher net data rates. A key enabler for PAS is the distribution matcher (DM), which maps uniform bits to symbols with desired properties (e.g., a good empirical distribution, low average power) in a reversible manner. Most work so far considered constant composition distribution matching (CCDM) [2], which has excellent performance but introduces additional complexity because of the arithmetic coding. In [3], the authors extend PAS to higher dimensions and suggest a shell mapper as DM to select low power sequences from a 4-dimensional (4D) hypercube consisting of the 4-fold Cartesian product of $M$-ary amplitude shift keying (ASK) constellations.

In this paper, we show experimentally that rate-adaptation with steps of 0.5 bits/QAM symbol (bpQs) can be achieved with the scheme of [3] and a simple look-up table. We compare this method to conventional PAS with CCDM and uniform constellations. The results are based on short-reach, unrepeated transmissions of 100 km, 140 km and 180 km.

## 2. Rate-Adaptation and Achievable Rates

### *2.1. Signaling Schemes*

We distinguish three signaling and receiver schemes. PAS-$n$D-1D is the conventional setting using a high, $n$-dimensional ($n$D) DM and 1D demapping. PAS-4D-4D uses a 4D DM and also demaps in 4D. A variant of the latter is PAS-4D-2D, where a sub-optimal and less complex demapping in 2D is used. The PAS-4D schemes employ 4D signaling with an $M$-ASK constellation in each dimension. For the binary labeling of the constellation points, we use an $m = \log_2(M^4)$ bit binary reflected Gray code (BRGC). Following the principle of quadrant shaping (QS) [3], $m_\text{Q} = m - 4$ bits determine a point in a quadrant and $m_\text{S} = 4$ bits represent the 4 sign bits. A rate-adaptation can be realized by a DM based on a look-up table with at most $2^{m_\text{Q}}$ entries, which selects a subset of $2^k$ low-energy points from the $2^{m_\text{Q}}$ points in a quadrant. The corresponding DM rate is $R_\text{dm} = k$ bits/4D-symbol. We refer to the resulting constellation as $\mathscr{X}$. The SE is $R_\text{tx} = R_\text{dm} + 4 \cdot \gamma$ bits/4D-symbol, where $\gamma$ is the fraction of sign bits per dimension that carry additional information bits. For the forward error correction (FEC) code rate $R_\text{c}$ we have $\gamma = 1 - (1 - R_\text{c}) \cdot \log_2(M)$. Its blocklength is usually an integer multiple of $m \cdot n$. By choosing $k$, a rate adaptation with a granularity of 1 bit/4D-symbol, i.e., 0.5 bpQs is possible so that any SE within the set $\{0.5, 1.0, 1.5, \ldots, 2\log_2(M) - 2\} + 2 \cdot \gamma$ bpQs can be realized with the *same* FEC overhead (OH). If the scheme is extended to $N$-dimensions ($N$D), a granularity of $1/(N/2)$ bpQs is possible. The exemplary transmission modes of this work target SEs of 3 bpQs, 4 bpQs and 5 bpQs and are summarized in Table 1a. We assume a *single* FEC code with OH = 23% ($R_\text{c} = 13/16$).

| Signaling mode | $R_{tx}$ [bpQs] | $\lvert\mathscr{X}\rvert$ | $R_{dm}$ [bpQs] | OH |
|---|---|---|---|---|
| PAS-4D-4D-4.5 | 5.0 | 8182 | 4.5 | 23% |
| PAS-4D-4D-3.5 | 4.0 | 2048 | 3.5 | 23% |
| PAS-4D-4D-2.5 | 3.0 | 512 | 2.5 | 23% |
| PAS-*n*D-1D-4.5 | 5.0 | 256 | 4.5 | 23% |
| PAS-*n*D-1D-3.5 | 4.0 | 256 | 3.5 | 23% |
| PAS-*n*D-1D-2.5 | 3.0 | 256 | 2.5 | 23% |
| 16-QAM uniform | 3.0 | 16 | – | 33% |
| 64-QAM uniform | 4.0 | 64 | – | 50% |
| 256-QAM uniform | 5.0 | 256 | – | 60% |

(a) Signaling modes.

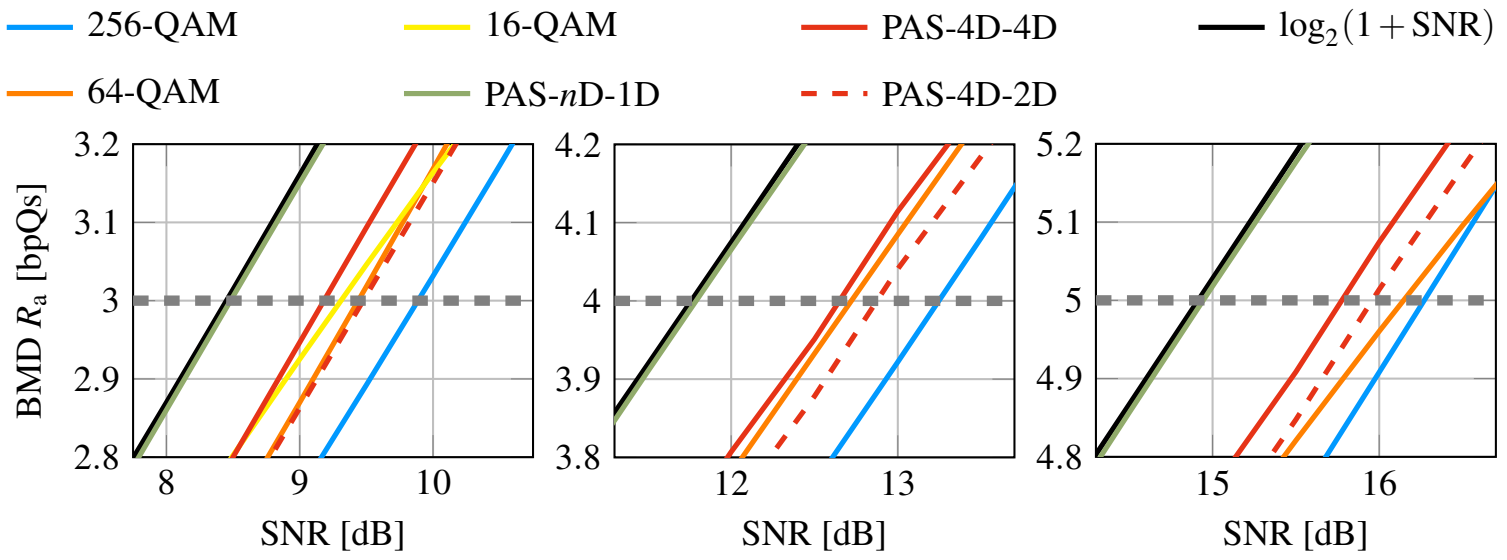

(b) $R_{tx} = 3$ bpQs (c) $R_{tx} = 4$ bpQs (d) $R_{tx} = 5$ bpQs

Fig. 1: Simulated achievable rates of several signaling strategies for the linear AWGN channel.

### 2.2. *Achievable Rates*

We denote the 4D transmit and receive vectors as $\boldsymbol{x} = (x_1,x_2,x_3,x_4) \in \mathscr{X}$ and $\boldsymbol{y} = (y_1,y_2,y_3,y_4) \in \mathbb{R}^4$. During the transmission experiment, we acquire sample sequences $\boldsymbol{x}^K = (\boldsymbol{x}_1,\boldsymbol{x}_2,\dots,\boldsymbol{x}_K)$ and $\boldsymbol{y}^K = (\boldsymbol{y}_1,\boldsymbol{y}_2,\dots,\boldsymbol{y}_K)$. An estimate of the achievable rate is [4]

$$R_a = \left[ \mathrm{H}(\boldsymbol{X}) - \frac{1}{K}\sum_{i=1}^{K} -\log_2\left(\frac{q(\boldsymbol{x}_i,\boldsymbol{y}_i)}{\sum_{\boldsymbol{a}\in\mathscr{X}} q(\boldsymbol{a},\boldsymbol{y}_i)}\right)\right]^+ \qquad \text{[bits/4D symbol]} \tag{1}$$

where $q(\boldsymbol{x},\boldsymbol{y})$ is the decoding metric for symbol-metric decoding (SMD) or bit-metric decoding (BMD) and $[\cdot]^+$ is short for $\max(0,\cdot)$. For PAS-4D-2D, the decoding metric may also implement a sub-optimal and less complex rule based on the respective 2D marginals.

The calculated achievable rates for BMD and the *linear additive white Gaussian noise (AWGN) channel* are shown in Fig. 1. We observe that the PAS-4D modes are superior to their uniform $\{16,64,256\}$-QAM counterparts for all three target SEs. The loss in power efficiency due to 2D demapping is at most 0.2 dB for BMD. The PAS-*n*D-1D modes virtually achieve Gaussian capacity.

## 3. Transmission Experiment

### 3.1. *Setup*

We experimentally investigate the 4D signaling scheme for a short-reach scenario and unrepeated transmission for 100 km, 140 km and 180 km, where the four dimensions are transmitted as two complex dimensions in two subsequent time slots. The CCDM for PAS-*n*D-1D operates in $n = 6000$ dimensions and the input distributions are taken from the Maxwell-Boltzmann family. The setup is shown in Fig. 2. The data symbols are interleaved with quadrature phase-shift keying (QPSK) pilots at a pilot rate of 10%. A frame alignment sequence is added at the beginning of the sequence and square root raised cosine (RRC) pulse shaping with a roll-off factor of 0.1 is applied [6]. We employ a wave-division multiplexing (WDM) setup with 5 external cavity lasers (ECLs, 10 kHz linewidth) on a 25 GHz grid. An arbitrary waveform generator (AWG, 20 GHz) drives the two IQ modulators. The four interferers (IQ mod 1) are combined with the central channel (IQ mod 2), a delay-and-add polarization emulator generates a dual-polarization signal and the channels are individually decorrelated. The transmission link consists of an erbium doped fiber amplifier (EDFA) followed by a variable optical attenuator (VOA) that sets the total power launched into the standard single-mode fiber (SSMF) of lengths 100 km, 140 km and 180 km. After transmission, the central channel is demodulated using a standard preamplified coherent receiver followed by a digital storage oscilloscope (DSO, 80 GSa/s and 33 GHz analog bandwidth). The receiver DSP is performed offline [6].

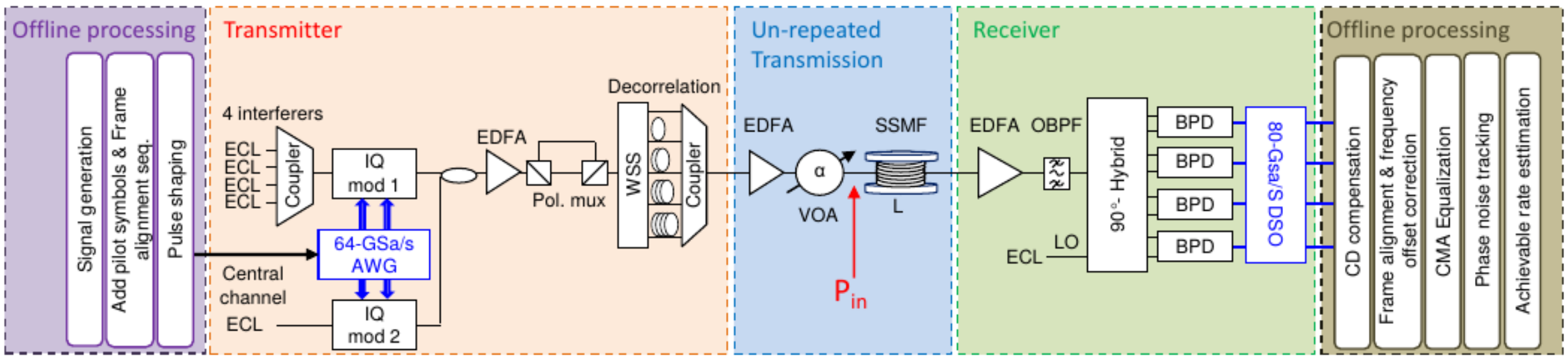

Fig. 2: Experimental setup of the transmission experiment.

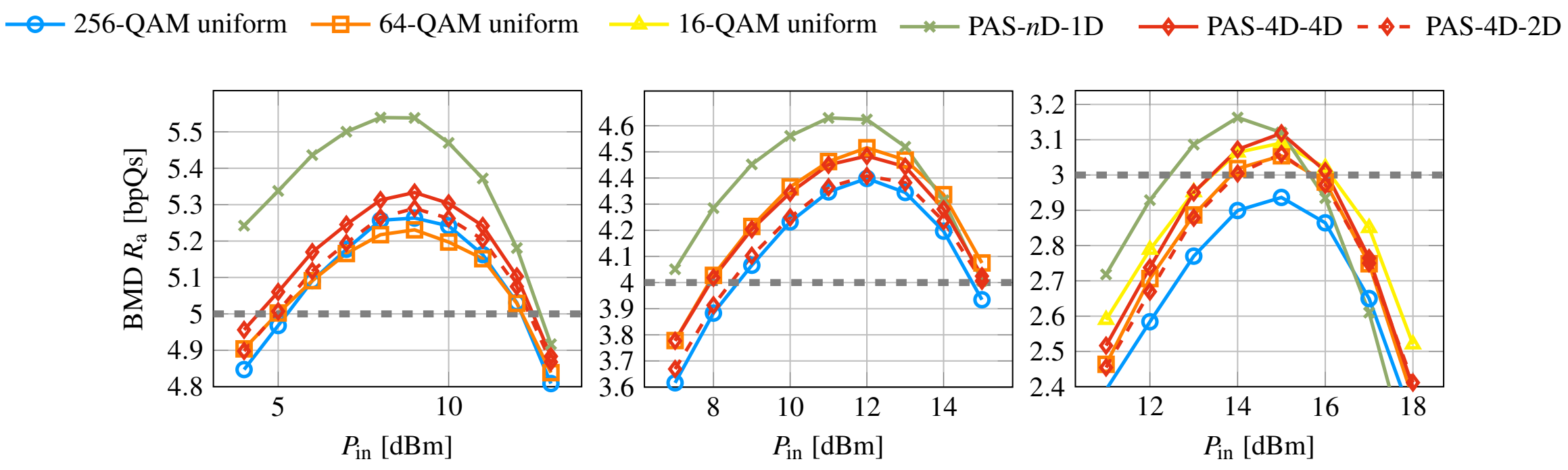

Fig. 3: Achievable rates of the considered signaling strategies for the transmission experiment.

### 3.2. *Results*

In back-to-back experiments, we validated that all modulation formats exhibit the same implementation penalty in the operating regime of interest to ensure a fair comparison. The achievable rates for the transmission experiments for link lengths of 100 km, 140 km and 180 km are shown in Fig. 3. To operate efficiently for all three distances (corresponding to optimal launch powers of 9 dBm, 12 dBm and 15 dBm) at the desired SEs (dashed gray line), the three different uniform constellations (16-, 64- and 256-QAM) would need to be operated with three different FEC codes (see Table 1a). The PAS modes allow a flexible operation with a single FEC. For increased transmission lengths, and therefore launch powers $P_{\text{in}}$, we observe that the PAS-$n$D-1D modes are penalized due to their increased higher order moments and their severe impact on the non-linear interference noise (NLIN) [5, 6]. At the respective optimal launch powers, the shaping gain of PAS-$n$D-1D is 0.2 bpQs, 0.15 bpQs, and 0.04 bpQs as compared to PAS-4D-4D. However, the PAS-4D DM is a simple look-up table with at most $2^k = 2^9 = 512$ entries. Moreover, 2D demapping (PAS-4D-2D) loses at most 0.1 bpQs for a target SE of 4 bpQs.

## 4. Conclusions

4D signal shaping and rate adaptation with a simple look-up table based DM allows a good trade-off between shaping gain and DM complexity. The suggested schemes are promising for transmission links where full fledged PAS is too complex or penalized because of NLIN. Future work should investigate the tradeoff between higher dimensional constellations with $N > 4$, rate flexibility, and demapping complexity.